\documentclass[12pt]{article}

\usepackage{latexsym, amsmath}
\newtheorem{theorem}{Theorem}
\newtheorem{lemma}{Lemma}
\newtheorem{corollary}{Corollary}

\newenvironment{proof}{{\it Proof:\/}}{\hfill $\Box$\\ }

\newcommand{\F}{\mbox{\bf  F}}

\title{Hard Problems of Algebraic Geometry Codes}

\date{}

\author{Qi Cheng\thanks{School of Computer Science,
the University of Oklahoma,
Norman, OK 73019, USA.
Email: {\tt qcheng@cs.ou.edu.}
This research is partially supported by NSF Career
Award CCR-0237845
}
}

\begin{document}

\maketitle

\begin{abstract}

The minimum distance is one of the most important combinatorial
characterizations of a code. The maximum likelihood decoding
problem is one of the most important algorithmic problems of a code.
While these problems
are known to be hard for general linear codes, the techniques used to
prove their hardness often rely on the construction of artificial
codes.
In general, much less is known about
the hardness of the specific classes of natural linear codes.
In this paper, we show
that both problems are
 NP-hard for algebraic geometry codes.
We achieve this by reducing a well-known NP-complete problem 
to these 
problems using a randomized algorithm. 
The family of codes in the reductions 
are based on elliptic curves. They have positive rates,
but the alphabet sizes are exponential 
in the block lengths.
\end{abstract}

\section{Introduction}

An $[n,k]_q$ linear error-correcting code is a linear 
subspace of a vector space $\F_q^n$, where $\F_q$ denotes the
finite field of $q$ elements, 
and $k$ denotes the dimension of the subspace. The
{\em Generator Matrix} for a linear code is a $k \times n$ 
matrix, with row rank $k$ which defines a linear mapping from $\F_q^k$ (called
the {\em message space}) to
$\F_q^n$. Therefore, the code $C$ is:
$$ C =\{ a G | a\in \F_q^k\}. $$
We call a vector in $C$ a codeword.
The most important codes include the Reed-Solomon codes,
the Reed-Muller codes,
the BCH codes and the algebraic geometry codes.

The {\em Hamming Distance} between two codewords $x$ and $y$,
is the weight (number of nonzero coordinates) of $x-y$. 
The minimum distance of a code is the minimum
Hamming distance between any two codewords.
If the code is linear,
then the vector $x-y$ is a codeword, and the minimum distance of the
code is equal to the minimum weight of any codeword.

Given a linear code as input, 
how hard is it to compute the minimum distance?
 This problem had been open for two decades
before it was finally solved by Vardy in 1997 \cite{Vardy97},
when he proved that the problem is NP-complete.
Interestingly, determining whether a code contains
a codeword of a given weight was known to be NP-complete much 
earlier \cite{BerlekampMc78}.
However,  if we know that the minimum distance of a code
is $d$, it merely implies that there is a codeword
of weight $d$, and for any $w<d$, there is no codeword
of weight $w$. It is not clear that for any $n \geq w>d$,
whether there exists a codeword of weight $w$ or not.
Thus there is no straight-forward  reduction from this problem
to the minimum distance problem.

Dumer et.al. \cite{DumerMi03}  studied how hard it
is to approximate the minimal distance of a linear code.
They showed that the minimum distance
of a linear code is not approximable  within
any constant factor in random polynomial time,
unless NP$=$RP.
The codes used in the work
of them and Vardy \cite{Vardy97} are
artificially designed. Their results exhibit that
it is hard to compute the minimum distance for the {\em general} linear codes,
but  say nothing specific about any of the well-studied
and widely-deployed codes.

To use a code in practice, one must have an efficient
decoding algorithm. Traditionally, {\em unique decoding algorithms}, 
which correct errors of weight at most half of the minimum distance
of a code,
have been investigated for natural classes of codes. The discovery of
such algorithms, which provide a means to correct errors,
enable the widespread application of error-correcting codes.
The {\em list decoding problem} can 
correct more errors and outputs a list of
codewords, any of which may be the intended message.
In the last decade,
spectacular success in the area of list decoding has been achieved,
its influence can be seen throughout theoretical computer science,
ranging from the approximation algorithm and the average case complexity,
to pseudorandomness and derandomization.
The ultimate goal, the {\em maximum likelihood decoding} problem, is
one of the central problems in algorithmic coding theory. For any
vector $y$ in $\F_q^n$, it asks for a codeword $x$ to minimize the
distance between $x$ and $y$. Given that a received word is equally
likely to contain an error in any position, codewords that are closest
to the received word (i.e. differ in fewer coordinates) are most
likely to encode the intended message.  This problem is proved to be
NP-hard for general linear codes \cite{BerlekampMc78}.  Proving
NP-hardness for the classes of useful codes is more difficult and
subtle.  The only result of this kind to date  is the result of
\cite{GuruswamiVa05} on the NP-completeness of maximum likelihood
decoding for Reed-Solomon codes. A related result by Cheng and Wan 
\cite{ChengWa04}
shows that decoding of Reed-Solomon codes at certain radius
is at least as hard as discrete logarithm problem over finite fields.

In this paper, we prove that the minimum distance problem
and the maximum likelihood decoding problem are NP-hard
for a natural class of codes, namely, the algebraic-geometry
codes. The algebraic geometry codes can be seen as 
a  generalization of the
Reed-Solomon codes. 
While the study of algebraic geometry codes began as a 
purely mathematical pursuit, an increased understanding of their
unique combinatorial properties 
promises that they 
will find real-world applications
in the foreseeable future.

In combinatorics, it is often hard to explicitly
construct an  object which is, in certain aspects, better than 
a random object.
A family of algebraic geometry codes 
is one of a few bright spots,
where we can  explicitly construct a code having more
codewords than a random code given the block length and the
minimum distance.
Moreover,  given proper representations, these codes  possess a
polynomial time list decoding algorithm \cite{GuruswamiSu99},
which corrects errors well beyond half of the minimum distance.
In contrast, a random code usually does not have a good decoding algorithm
due to its lack of algebraic structure.

Proving the NP-hardness of the maximum likelihood
decoding of algebraic geometry codes (MLDAGC) 
answers the most important question
about the decodability of this class of codes.
Proving the NP-hardness of the
minimum distance problem for algebraic geometry codes (MDPAGC) is also 
well motivated.
The designed distance, which is a lower bound of
the minimum distance,  can be easily obtained from 
the description of the codes. Less attention is paid to the problem
of computing the exact minimum distance.

Also, the minimum distance problem for general linear codes
defied solution for so long time, one would imagine that
the problem for  codes with algebraic structures is more subtle.
If a code has a good list decoding algorithm,
while at the same time computing its minimum distance is hard,
then  we cannot easily find
a center of a Hamming ball with
the list decodable radius that contains two codewords at 
the minimum distance from each other. This illustrates deep structural
information about the code which may uncover properties of the
code that we have not yet realized.

A nice surprise about our proofs is its conceptual simplicity.
We use the subset sum problem directly, thus all of
the results on the preprocessing subset sum problem 
can be readily carried over to the algebraic geometry codes.
However our reductions are
randomized, which we would prefer to avoid. 
 The need for randomization seems to occur in places
where we  deal
with number theory and primes. 
In \cite{Vardy97} and \cite{GuruswamiVa05}, 
an irreducible polynomial over $\F_2$ 
is needed. Even though there is no polynomial time algorithm
which finds an irreducible polynomial over a finite field
of a given degree,
there does exist a deterministic algorithm
which finds an irreducible polynomial of a given degree
over finite fields of fixed number of elements \cite{Shoup90}.
This explains why the reduction in \cite{Vardy97} and \cite{GuruswamiVa05}
is deterministic.

Our reduction always maps a ``Yes'' instance to a ``Yes'' instance,
and maps a ``No'' instance to a ``No'' instance
in {\em expected polynomial time}. The reductions in \cite{DumerMi03}
is a {\em reverse unfaithful random reduction}, which always maps a
``No'' instance to a ``No'' instance, but with a small probability,
maps  a ``Yes'' instance to a ``No'' instance.

The minimum distance problem, and the maximum likelihood
decoding problem, correspond to 
the shortest vector problem and the closest vector problem
in integral lattices. 
These problems have received a lot of
attentions recently \cite{Ajtai98, Khot04}. 
The attempts to
find a reduction from the minimum distance problem
of linear codes to the shortest vector problem 
of lattices
have failed so far.

\section{Elliptic curves}

The Reed-Solomon code of block length $n$ and dimension $k$
is obtained by evaluating polynomials of degree $k-1$ at
a set of
$n$ many elements in a finite field.
For a linear $[n,k]_q$ code, the Singleton bound
asserts that $d \leq n -k +1 $.
The Reed-Solomon codes are optimal, in that they satisfy the
Singleton bound with equality. 
It is trivial to read the minimum distance of Reed-Solomon codes
from the block length and the dimension.

The algebraic geometry codes are natural generalizations of
the Reed-Solomon codes.
Let $K$ be a function field over a finite field $\F$.
Let $A_1, A_2, \cdots, A_n, B_1, B_2, \cdots, B_m$
be $\F$-rational places.
Let $a_1, a_2, \cdots, a_n, b_1, b_2, \cdots, b_m$
be positive integers.
Given a divisor $A = \sum_{i =1}^n a_i A_i - \sum_{i =1}^m b_i B_i$,
define $L(A)$ to be the set of functions, each has poles only at $A_1,
A_2, \cdots, A_n$ with multiplicities at most $a_1, a_2, \cdots, a_n$
respectively, has zeros at $B_1, B_2, \cdots, B_m$ with
multiplicities at least $b_1, b_2, \cdots, b_m$ respectively.
The functions in $L(A)$ form  a linear space over the field $\F$.
It has dimension no less than $deg(A) - g +1$,
where $g$ is the genus of the function field, and
$deg(A) = \sum_{i =1}^n a_i  - \sum_{i =1}^m b_i $.
For the divisor $A$, we can construct
a linear code, 
whose codewords are  obtained by evaluating the functions
in $L(A)$
at rational places $P_1, P_2, \cdots, P_n$,
where $\{P_1, P_2, \cdots, P_n \} \cap \{
A_1, A_2, \cdots, A_n, B_1, B_2, \cdots, B_m\} = \emptyset$.

To prove that computing minimum
distances of algebraic geometry codes is NP-hard, we use
codes defined by curves of genus one, i.e.,  elliptic curves.
we first review some facts about elliptic curves.
An elliptic curve is a smooth cubic curve.
Let $\F$ be a field. If the characteristic of $\F$
is neither $2$ nor $3$,
we may assume that an elliptic curve is given
by an equation 
$$ y^2 = x^3 + ax +b, \hspace{0.5in} a,b\in \F.$$
The discriminant of this curve is defined as $-16(4a^3 + 27b^2)$.
It is essentially  the
discriminant of the polynomial $x^3 + ax + b$.
It should be non-zero for the curve is smooth.
For detailed information about elliptic curves,
we refer the reader to Silverman's book \cite{Silverman86}.
The
set of $\F$-rational
points on the elliptic curve consists of the solution set over $\F$
of the
equation plus a point at infinity, denoted by $O$. 
These points form an
abelian group with the infinity point as the identity.  We use $E(\F)$
to denote the group. From now on, let $\F$ be the finite field $\F_q$.
The following properties of elliptic curves
are relevant to our result.

\begin{enumerate}
\item Let $P_1, P_2, \cdots, P_n, P$ be elements in $E(\F_q)$.
If $m_1 P_1 + m_2 P_2 + \cdots + m_n P_n = P$,
where $m_i$, $1\leq  i \leq n$, are positive integers,
then
there is a function having zeros at $P_1, P_2, \cdots, P_n$,  with
multiplies $m_1, m_2, \cdots, m_n$ respectively, 
a pole at $P$ with multiplies $1$
and a pole at $O$ with multiplies $m_1 + m_2 + \cdots + m_n -1$.
We can compute the function in time
polynomial in $m_1+ m_2+\cdots+m_n$ and $\log q$ \cite{HuangIe94}.
\item For a given divisor $A$, we can in polynomial time
compute a basis of $L(A)$. 
In particular, since $(x)_{\infty} = 2O$,
$(y)_{\infty} = 3O$, and consequently, $(x^i)_{\infty} = 2iO$,  
$(x^{i-1} y)_{\infty} = (2i+1)O$,  we can compute a basis
for $L(\alpha O)$ quickly, and it contains only monomials.
\item If $deg(A)\geq 1$, then dimension of $L(A)$ is $deg(A)$.
\item Let $p \equiv 2 \pmod{3} $ be a prime.
The curve $y^2 =x^3 +1$ is a supersingular elliptic curve over $\F_p$.
The group $E(\F_p)$ contains $p+1$ elements and it is cyclic.
\end{enumerate}

\begin{lemma}\label{curveconstruction}
For any prime $q > 3$, we can in randomized polynomial time
find another prime $p = O(q^2)$ and
construct an elliptic curve $E/\F_p$ and a point $G\in E(\F_p)$
such that the  $G$ has order $q$.
\end{lemma}

\begin{proof}
Find another prime $p $ such that $ p \equiv -1 \pmod{q}$ and
$p \equiv 2 \pmod{3}$. This can be done easily if randomness is
allowed.
We can first solve the system of congruences using
the Chinese Remainder Theorem. If the solution is $ p = a \pmod{3q}$,
we select a random number $1\leq x\leq q$, and test whether $ a + 3qx$
is prime or not. By the Siegel-Walfisz theorem concerning
the density of primes in arithmetic progression,
the probability that we get a prime
is at least  $1 / \log^{O(1)} 3q$. Set $p=a+3qx$ if we find a prime.

Consider the curve $E: y^2 =x^3 + 1$ over $\F_p$. It is supersingular
hence $E(\F_p)$ is a cyclic group with order $p+1$.
We try to find a point $P$ in the group such that ${p+1 \over q} P \not= O$.
Since the group is cyclic, the number of points $P$ such that 
${p+1 \over q} P = O$ is ${p+1 \over q}$,
so there is an overwhelming chance of success.
Once we find a $P$ satisfying
${p+1 \over q} P \not= O$, set $G = {p+1 \over q} P$. 
It is easy to verify that $G \in E(\F_p)$ is a point with order
$q$. 
\end{proof}

The curve we construct is supersingular, therefore it is not suitable
for elliptic curve cryptosystems if 
$p$ is small, since the discrete logarithm problem on those elliptic curves
can be reduced to the discrete logarithm problem in $\F_{p^2}$.
For practical purposes, there is an efficient method  
based on the theory of complex multiplication to
construct a nonsupersingular curve of  a given order, 
but 
it seems hard to prove the performance in theory.

In the proof, we need randomness  to find a large order point
on an elliptic curve. To deterministically
find any point on an elliptic curve is still an
open problem, even though 
an efficient and simple Las Vegas algorithm exists.

\section{The NP-hardness proof of the MDPAGC}

We reduce the following well known subset sum problem
to the problem of computing minimum distances 
of algebraic geometry codes.

\begin{description}
\item[Instance:] A set of $n$  positive integers 
$A = \{a_1, a_2, a_3, \cdots, a_n\}$, a positive integer $b$ and 
a positive integer $k < n$.
\item[Question:] Is there a nonempty subset $\{ a_{i_1}, a_{i_2}, \cdots,
a_{i_k} \} \subseteq A$ of cardinality $k$ such that 
$$ a_{i_1} + a_{i_2} + \cdots + a_{i_k} = b. $$ 
\end{description}

First we prove a slight variety of the problem is also NP-hard.

\begin{lemma}
The following problem ({\em prime field subset sum problem}) is NP-hard:
\begin{description}
\item[Instance:] A prime $q$, a set of $n$  positive integers 
$A = \{a_1, a_2, a_3, \cdots, a_n\} $, an integer $b$ and 
a positive integer $k < n$.
\item[Question:] Is there a nonempty subset $\{ a_{i_1}, a_{i_2}, \cdots,
a_{i_k} \} \subseteq A $ of cardinality $k$ such that 
$$ a_{i_1} + a_{i_2} + \cdots + a_{i_k} = b \pmod{q}. $$ 
\end{description}
\end{lemma}

To prove the lemma, we simply reduce the
subset sum problem to it by finding a prime bigger than
$ a_1 + a_2 + a_3 + \cdots + a_n  +b$ in an instance of
the subset sum problem.
It is interesting to note that it seems hard to
prove the NP-completeness under the polynomial time 
Karp reduction, since such a reduction would give
rise to a deterministic algorithm to find 
a prime bigger than a given number,
but no such an algorithm is known. The problem was listed as
open in \cite{Adleman94}. Derandomizing the algorithm
is very interesting, given that 
a deterministic polynomial
time primality testing algorithms was discovered recently
\cite{AgrawalKa02}.

\begin{theorem}
Given a
instance of the prime field subset sum problem, 
we can in  randomized polynomial time, construct an algebraic geometry
code $[n,k]_p$ with $p = O(q^2)$ 
such that if the answer to the prime field subset sum problem is ``YES'',
then the code has minimum distance $n-k$. If the answer to the
prime field subset sum problem is ``NO'', then the code 
has minimum distance $n -k +1$.
\end{theorem}

\begin{proof}
Given an instance of the prime field subset sum problem,
by Lemma~\ref{curveconstruction}, we can construct 
an elliptic curve $E$ over $\F_p$, $p= O(q^2)$ , with a point $G$ of order $q$.
Let $Q = bG$.
Now consider an algebraic geometry codes generated by
evaluating functions in $L(Q + (k-1)O)$ at
$$P_1 = a_1 G, P_2 = a_2 G, \cdots, P_n = a_n G.$$
By the Singleton bound, we know that
the minimum distance is at most $n-k+1$.
This code has designed distance $n-k$, thus
the minimum distance is at least $n-k$. 
Let $f_1,f_2, \cdots, f_k$ be a basis of $L(Q+(k-1)O)$, 
the generator matrix of the code is
$$
\begin{pmatrix}
f_1 (P_1) & f_1(P_2) & \dots & f_1(P_n)\\
f_2 (P_1) & f_2(P_2) & \dots & f_2(P_n)\\
\hdotsfor{4}\\
f_k (P_1) & f_k (P_2) & \dots & f_k(P_n)
\end{pmatrix}
$$

If there exists a subset $\{ a_{i_1}, a_{i_2}, \cdots,
a_{i_k} \} \subseteq \{a_1, a_2, \cdots, a_n\} $ such that 
$ a_{i_1} + a_{i_2} + \cdots + a_{i_k} = b \pmod{q}, $
then $P_{i_1} + P_{i_2} + \cdots + P_{i_k} = Q$ in $E(\F_p)$.
Thus there exists a function $f$ having zeros at $P_{i_1}$,
$P_{i_2}$, $ \cdots, P_{i_k}$ with single multiplicity,
a pole at $Q$ with single multiplicity, and a pole at $O$
with multiplicity $k-1$. We have $f\in L(Q + (k-1)O)$.
Such a function is unique up to a constant factor.
The codeword corresponding to $f$ has weight $n-k$,
because it has $k$ zeros in $\{ P_1, P_2. \cdots, P_n\}$.

In the other direction, if the minimum weight of the codewords
is $n-k$, there exists a function $f \in L(Q + (k-1)O)$ whose has zeros at
$k$ many points in $P_1, P_2, \cdots, P_n$.
Denote them by $P_{i_1}, P_{i_2}, \cdots, P_{i_k}$.
Since it can  have no more than 
$k$ poles, counting multiplicities, 
it must have exactly $k$ zeros, and all the zeros have
single multiplicity.
Thus it must have $k$ poles as well. 
It has a pole at $Q$ with multiplicity $1$ and
a pole at $O$ with multiplicity $k-1$. That is to say
$(f) = P_{i_1} + P_{i_2} + \cdots + P_{i_k} -  Q - (k-1) O$.
Hence in $E(\F_p)$ 
$$P_{i_1} + P_{i_2} + \cdots + P_{i_k} =  Q.$$
We have
$$  a_{i_1}G + a_{i_2}G + \cdots + a_{i_k}G = b G. $$  
It implies that $a_{i_1} + a_{i_2} + \cdots + a_{i_k} = b \pmod{q}$.

\end{proof}

The reductions in the proofs are randomized.
We need to use randomness to find a prime of certain size
and a point on an elliptic curve of the prime order.
Once we find such a prime or point, we can
provide a proof of the primality or the order.
On the contrary, in Dumer et.al.'s work \cite{DumerMi03},
they need randomness to locate a good center,
for a Hamming ball of certain radius containing many codewords.
Even though with a high probability,
a random received word qualifies,
no proof of this fact can be provided.

\begin{corollary}
If there is a  polynomial time Las Vegas algorithm to compute the minimum
distance of an algebraic geometry code, then $NP\subseteq ZPP$.
If there is a  polynomial time randomized algorithm to compute
the minimum distance of an algebraic geometry code,
then $NP\subseteq RP$.
\end{corollary}

\begin{corollary}
Deciding whether an algebraic geometry code is
maximum distance separable is NP-hard.
\end{corollary}

We can also use one point divisor codes by
reducing the following problem to MDPAGC.
The detail will be left in the full paper.
\begin{description}
\item[Instance:] A set of $n$ integers $\{a_1, a_2, \cdots, a_n\}$ 
and $k$, a prime $q$.
\item[Question:] Are there $k$ integers $a_{i_1}, a_{i_2},
\cdots, a_{i_k}$ such that 
$$a_{i_1} + a_{i_2}  + \cdots + a_{i_k} \equiv 0 \pmod{q}$$
\end{description}



\section{A time complexity lower bound for computing the minimum distance}

For the above analysis, it is easy to see that
we can in time $2^n (\log q)^{O(1)}$ compute
the minimum distance of an elliptic code in $[n,k]_q$.
Does there exist a better algorithm? 
If a problem is NP-hard, we do not expect to find an algorithm solving
it in polynomial time, no even in subexponential time. 
However, for NP-hard problems, sometimes we 
can find exponential algorithms beating the trivial
exhaustive search. What can we do in the case of
the minimum distance problem of algebraic geometry codes?
We can ask the same question for
 general linear
codes as well: can we compute the minimum distance 
in time $2^{cn} (\log q)^{O(1)}$
for some small $c$?

Ajtai et.al. \cite{AjtaiKu01} have studied the problem.
They proposed an algorithm
that solves the problem in time $2^{O(n)}$
if the field size is bounded by a polynomial in $n$.
The exact constant hidden in big-O is not calculated
in their paper.

The elliptic curve discrete logarithm problem (ECDLP) 
is to compute $l$ such that $Q =  lP$, given $P,Q\in E(\F_q)$.
It is obviously
an NP-easy problem, and is not believed to
be NP-hard. This is for sure a randomized
polynomial time reduction from the ECDLP 
to any NP-hard problem, including the minimum distance
problem of an algebraic geometry code.
In this section, we present a succinct reduction.
We reduce ECDLP over $\F_q$ to the problem
of computing the minimum distance of algebraic
codes in $[n, k]_q$, where $n \leq \lfloor \log q\rfloor$.

It is assumed in the elliptic curve cryptography that
there is no algorithm which runs in time $q^c$ for $c <  1/2$
to solve ECDLP in $\F_q$.
Under the assumption, we  have a
lower bound on the time complexity of computing
the minimum distance of linear codes.

\begin{theorem}
For any constant $c>0$,
if there is an algorithm which in time $2^{c n}(\log q)^{O(1)}$
computes the minimum distance of a linear code $[n,k]_q$,
then the ECDLP over $\F_q$ can be solved in time $q^c$.
\end{theorem}

\begin{proof}

Suppose that we need to compute the discrete logarithm of
$Q$ base $P$ on elliptic curve
$E(\F_q)$. W.l.o.g, we assume that $P$ has a prime order 
$p $. Note that we must have $p \leq q +1 -2\sqrt{q}$.

Denote the largest even number which is not bigger than
$\lfloor \log p \rfloor$ by $n$.
Randomly select a positive integer $r < p$, 
computer $R = rQ$. With probability ${ n \choose n/2}/2^n > 1/n^{O(1)}$,
the discrete logarithm of $R $ is an integer, when written in binary,
has exactly $n/2$ ones and $n/2$ zeros.

Now consider the code $C$ generated by evaluating functions in
$L(R + (n/2-1) O)$ at $P_0 = P, P_1 = 2P, P_2 = 2^2P, \cdots, 
P_{n-1} = 2^{n-1} P$.
By the similar reasoning, the minimum distance of the code
is $n/2$ iff $R$ can be written as a sum of $n/2$ points
from $P_0, P_1, \cdots, P_{n-1}$. Denote the set of these $n/2$ points
by $D$.
Let $C_i$ be the code  generated by evaluating functions in
$L(R + (n/2-1) O)$ at $P_0, P_1, \cdots, P_{i-1}, P_{i+1},
\cdots, P_{n-1}$.
We can find $D$
by asking the question where
the minimum distance of $C_i$, for $1\leq i\leq n$,
is $n/2$.
Basically,  $P_i\in D$ iff the answer for $C_i$ is ``No''.
We  solve the discrete logarithm problem immediately
after we get $D$.
\end{proof}

\section{The maximum likelihood decoding for AG-codes is NP-hard}

The dimension of linear space $L( (k-1) O) $ over $\F_q$ 
is $k-1$ for an elliptic curve. The dimension of linear space
$L(Q + (k-1) O)$, $Q \not= O$, is $k $.  Let $f_1, f_2, \cdots, f_{k-1}$
be a basis for $L((k-1) O) $, and  $f'$ be a function in 
$  L(Q + (k-1) O) - L((k-1) O) $.
Then $f_1, f_2, \cdots, f_{k-1}$
and $f'$ form a basis for $L(Q + (k-1) O)$.
It is fairly easy to find an $f'$. We can simply
pick one point $Q' \not\in \{Q, O \}$, compute $Q'' = Q - Q'$.
Let $l_1$ be the line passing $Q'$ and $Q''$, let $l_2$ be the line
passing $Q$ and $-Q$. We then set $f' = l_1/l_2$.

\begin{lemma}\label{distancelemma}
Consider the code generated by evaluating functions
in $L((k-1) O) $ at $P_1, P_2, \cdots, P_n$.
Suppose the received word is $R = (f'(P_1), f'(P_2), \cdots, f'(P_n))$.
Then 
\begin{enumerate}
\item the distance from $R$ to the code is either $ n - k +1$ or $ n-k $
\item the distance from $R$ to the code is $n- k $
iff there is a subset $P_{i_1}, \cdots, P_{i_{k}}$
of $P_1, P_2, \cdots, P_n$ such that
 $$  P_{i_1}+ P_{i_2} +  \cdots + P_{i_{k}} = Q  $$ 
\end{enumerate}
\end{lemma}

\begin{proof}

It is clear that  $R$ is not a codeword, since
if $f'\in L(Q+(k-1)O)$ takes the same values  as a function in $L((k-1) O)$
at $n$ distinct points, it must be equal to the function,
but $f'$ has a pole at $Q$.

If the distance is   less than $n-k$,
it means that there is a function 
$f \in L( (k-1) O)$ such that $f'-f$ has more than $k$ distinct zeros
in $\{P_1, P_2, \cdots, P_n \}$.
But $f'-f\in L(Q + (k-1)O)$, it has at most $k$ poles. A contradiction.

If the distance from $R$ to the code is $n-k$,
there is a function 
$f \in L( (k-1) O)$ such that $f'-f$ has $k$ distinct zeros.
Let them be $P_{i_1}, \cdots, P_{i_{k}} $.
The function $f'-f$  
must have a pole at $Q$ with multiplicity $1$ and a pole at $O$ with
multiplicity $k-1$. Therefore,
we have $ (f'-f) = P_{i_1} + \cdots + P_{i_{k}} - Q - (k-1) O$ and 
in $E(\F_p)$
   $$  P_{i_1} + \cdots + P_{i_{k}} = Q.  $$

In the other direction, if there is a 
subset $P_{i_1}, \cdots, P_{i_{k}}$
of $P_1, P_2, \cdots, P_n$ such that
 $$  P_{i_1}+ P_{i_2} +  \cdots + P_{i_{k}} = Q  $$ 
This implies that there is a function $g $ such that
$$(g) =  P_{i_1} + \cdots + P_{i_{k}} - Q - (k-1) O.$$
It is clear that $g\in L(Q + (k-1)O)$, thus
$g = f + a f'  $, where
$f \in L( (k-1)O)$  and $a\not= 0$.
The vector $R$ is at distance $n-k$ away from the codeword
obtained by evaluating the function
$-f/a $ at $P_1, P_2, \cdots, P_n$.

To prove that the distance is at most $n-k+1$, 
compute $P' = Q - P_1 - P_2 - \cdots - P_{k-1}$.
If $P' \in \{ P_{k}, P_{k+1}, \cdots, P_n \}$,
then we have shown that the distance 
from $R$ to the code is $n-k$. Assume that
it is not the case. There exists a function $g'$ such that
$$ (g') = P_{i_1} + \cdots + P_{i_{k-1}}+P' - Q - (k-1) O. $$
Since $g' \in L(Q + (k-1)O)$, we have that $g' = a f' + f$
for some $f\in L((k-1)O)$ and $a\in \F_q^*$. 
This shows that the distance from $R$ to
the code is not longer than $n-k+1$.
\end{proof}

\begin{theorem}
Given a received vector,
computing the distance from the vector 
to an elliptic code is NP-hard.
Therefore, the maximum likelihood decoding problem for
algebraic geometry codes is NP-hard.
\end{theorem}

\begin{proof}
Given an instance of the prime field subset sum problem,
we construct 
an elliptic curve $E$ over $\F_p$, $p= O(q^2)$ , with a point $G$ of order $q$.
Let $Q = bG$, and let $f'$ be a function in $L(Q+(k-1)O) - L((k-1)O)$.
Now consider an algebraic geometry code generated by
evaluating functions in $L( (k-1)O)$ at
$P_1 = a_1 G, P_2 = a_2 G, \cdots, P_n = a_n G$.
According to Lemma~\ref{distancelemma},
the answer to the prime field subset sum instance 
is ``Yes'', iff the distance
from $R = (f'(P_1), f'(P_2), \cdots, f'(P_n))$ to the code is $n-k$.

\end{proof}

Applying the result about the preprocessing subset sum problem 
\cite{Lobstein90}, we get

\begin{corollary}
There is a sequence of algebraic geometry codes $C_1, C_2, \cdots,
C_i, \cdots$, where $C_i \in [i,k]_{q_i}$,
such that the existence of polynomial size circuits
which solve their maximum  likelihood decoding problems implies that
$NP\subseteq P/poly$.
\end{corollary}

\section{Concluding remarks}

In this paper, we prove that computing  minimum distances
and the maximum likelihood decoding 
are NP-hard for algebraic geometry codes.
Our results rule out the possibility
of polynomial time solutions for these two problems, unless $NP=ZPP$.

The Reed-Solomon codes can be thought of as a special
case of algebraic geometry codes, in which
we use the rational function field. 
Let $O$ be the infinity point on the projective line.
The functions $1, x, x^2, \cdots, x^k$ form
a basis for $L(kO)$. In \cite{ChengWa04}, the authors study Hamming balls 
centered at the vectors $(r(x)/h(x))_{x\in \F_q}$, where $r$ and
$h$ are polynomials in order  
to prove that the bounded distance decoding
for the Reed-Solomon codes is hard.
The function $f(x)/h(x)$ has poles at point other than $O$.
Some results in  \cite{GuruswamiRu05} follow a similar line.
In the proof of Lemma \ref{distancelemma}, 
we use $f'$ to generate a received word, it has poles
at a place other than $O$.
We suspect that further exploration of this connection between rational
functions with a different pole 
and decoding  problems would prove fruitful.

Our results use algebraic geometry codes based on elliptic curves.
In many ways, the elliptic codes are very similar to
the Reed-Solomon codes.
Intuitively we expect that the decoding problem for elliptic codes
is the easiest among all algebraic geometry codes.
We leave it as an open problem to
prove that both problems are NP-hard for codes based on curves
of any fixed genus.

The most interesting family of algebraic geometry codes 
has a fixed alphabet.
The codes in our results have alphabets of exponential size.
Nonetheless, we observe that
all the known decoding algorithms for algebraic geometry
codes are not sensible to the size
of the alphabets. Our results indicate that if 
a polynomial time maximum likelihood decoding
algorithm for  algebraic geometry codes 
does exist, it can only work for codes with a small alphabet size.
We conjecture that the maximum likelihood decoding is NP-hard
even for a family of
algebraic geometry codes with a fixed alphabet,
and leave it as an open problem.

\section*{Acknowledgments}
We thank Daqing Wan and Elizabeth Murray for helpful discussions.

\end{document}